\documentclass[]{fairmeta}
\usepackage[T1]{fontenc}
\usepackage{graphicx}
\usepackage{amsmath,amssymb,mathtools}
\usepackage{booktabs}
\usepackage{multirow}
\usepackage{algorithm}
\usepackage{algpseudocode}
\usepackage{subcaption} 
\usepackage{tikz}
\usepackage{quantikz}
\usepackage{hyperref}
\usepackage{tikz}
\usepackage{quantikz}   
\usepackage{pgfplots}   
\pgfplotsset{compat=1.18}
\usepackage{subcaption} 
\usepackage{amsthm}
\newtheorem{theorem}{Theorem}

\title{QuCoWE: Quantum Contrastive Word Embeddings with Variational Circuits for Near‑Term Quantum Devices}

\author[1\dagger,3]{Rabimba Karanjai}
\author[3]{Hemanth Hegadehalli Madhavarao}
\author[2]{Lei Xu}
\author[1]{Weidong Shi}

\affiliation[1]{University Of Houston}
\affiliation[2]{{Kent State University}}
\affiliation[3]{PayPal AI Lab}
\contribution[\dagger]{rkaranjai@uh.edu}

\abstract{
    We present \textbf{QuCoWE}, a framework that learns \emph{quantum‑native} word embeddings by training parameterized quantum circuits (PQCs) with a contrastive skip‑gram objective. Words are mapped to shallow, hardware‑efficient circuits with data re‑uploading and controlled ring entanglement; similarity is computed via quantum state overlap (fidelity) and passed through a \emph{logit‑fidelity} scoring head that recovers the shifted PMI semantics of SGNS/Noise‑Contrastive Estimation. We introduce an \emph{entanglement budget} regularizer based on single‑qubit purities to keep circuits trainable (mitigating barren plateaus with local costs and bounded entanglement), and we give a noise analysis for depolarizing and readout errors with error‑mitigation hooks (zero‑noise extrapolation, randomized compiling). On Text8/WikiText‑2 pretraining, QuCoWE reaches competitive intrinsic (WordSim‑353, SimLex‑999) and extrinsic (SST‑2, TREC‑6) performance versus 50–100d classical baselines while using fewer learned parameters per token via compact PQCs. We ablate qubit counts, re‑uploading depth, and entanglement patterns, and report training behavior under noise. Our results suggest that shallow PQCs with calibrated scoring and entanglement control are a viable path to distributional semantics on near‑term devices, and they connect classical PMI objectives with quantum fidelity via NCE.
}

\date{\today} 

\begin{document}
\maketitle

\section{Introduction}

Word embeddings form the foundation of modern natural language processing, encoding semantic relationships through distributional statistics \cite{mikolov2013distributed,pennington2014glove}. Classical approaches like Word2Vec and GloVe represent words as real-valued vectors optimized through contrastive objectives or matrix factorization. However, these representations face fundamental limitations in capturing complex semantic phenomena such as polysemy, compositionality, and graded entailment relationships.

Quantum computing offers a radically different representational paradigm through complex amplitudes, superposition, and entanglement. Recent advances in quantum natural language processing (QNLP) have explored quantum-inspired models \cite{coecke2020foundations}, but most rely on classical pre-training or simplified quantum simulations. The emergence of Noisy Intermediate-Scale Quantum (NISQ) devices \cite{preskill2018quantum} enables genuine quantum computation, albeit with significant noise and limited coherence.

This work addresses a fundamental question: \emph{Can we learn quantum-native word embeddings directly from text corpora that leverage unique quantum properties while remaining trainable on near-term devices?} We propose QuCoWE, a framework that represents words as parameterized quantum states trained through contrastive learning. Our approach bridges classical distributional semantics with quantum information theory, establishing theoretical connections between quantum fidelity and pointwise mutual information.

\subsection{Motivations and Challenges}

The motivation for quantum word embeddings stems from three key observations about the potential advantages of quantum representations. First, the representational capacity of quantum systems is fundamentally different from classical vectors. A quantum state on $n$ qubits requires $2^n$ complex amplitudes, offering exponential representational capacity that enables encoding rich semantic structures in compact representations. Second, quantum mechanics provides natural composition operations through tensor products and partial traces that align remarkably well with linguistic compositionality principles \cite{coecke2020foundations}. Third, quantum overlap captures both magnitude and phase relationships through fidelity measurements, potentially distinguishing semantic nuances that remain invisible to classical cosine similarity metrics.

However, realizing these theoretical benefits faces significant practical challenges that must be carefully addressed. The most severe obstacle is the barren plateau phenomenon, where random parameterized circuits suffer from exponentially vanishing gradients \cite{McClean2018NatComm}, rendering standard training procedures infeasible. Additionally, NISQ devices exhibit high error rates that can overwhelm the delicate quantum states encoding semantic information. Hardware constraints further complicate implementation by restricting available entanglement patterns and limiting circuit depth. Finally, bridging the gap between classical NLP objectives and quantum optimization requires careful architectural and algorithmic design to ensure meaningful learning.

To address these challenges while demonstrating the viability of quantum-native word embeddings, we study whether shallow, hardware-efficient PQCs can learn distributional semantics directly. Our contributions are:

\begin{itemize}
\item \textbf{Architecture.} A shallow data-reuploading ansatz with ring entanglement, designed for NISQ hardware \cite{PerezSalinas2020Quantum}.
\item \textbf{Logit-fidelity head.} A calibrated mapping aligning quantum overlap with shifted PMI under SGNS/NCE \cite{LevyGoldberg2014NIPS,GutmannHyvarinen2012JMLR}.
\item \textbf{Entanglement budget.} A purity-based regularizer that favors trainability consistent with local-cost barren-plateau theory \cite{Cerezo2021NatComm}.
\item \textbf{Noise analysis.} Closed-form effect of depolarizing noise on overlap, plus error-mitigation hooks (ZNE, randomized compiling) \cite{Temme2017PRL,MeyerWallach2002JMP}.
\end{itemize}

\textbf{Scope and Claims:} This work pursues NISQ-realistic resource efficiency rather than claiming unconditional quantum speedup. Concretely, QuCoWE matches the performance of classical 50--100 dimensional baselines on both intrinsic word similarity benchmarks and extrinsic text classification tasks, while using fewer learned parameters per token and demonstrating favorable sample efficiency. We restrict our current investigation to word-level embeddings, discussing extensions to sentence-level composition as important future work.
\section{Related Work}

\subsection{Classical Word Embeddings}

Distributional semantics hypothesizes that words appearing in similar contexts share meaning \cite{harris1954distributional}. Word2Vec \cite{Mikolov2013NIPS} operationalizes this principle through skip-gram with negative sampling (SGNS), optimizing:
\begin{equation}
\mathcal{L}_{\text{SGNS}} = -\sum_{(w,c) \in D^+} \log \sigma(v_w \cdot v_c) - \sum_{(w,n) \in D^-} \log \sigma(-v_w \cdot v_n)
\end{equation}

Levy and Goldberg \cite{LevyGoldberg2014NIPS} proved that SGNS implicitly factorizes a shifted PMI matrix:
\begin{equation}
v_w \cdot v_c \approx \text{PMI}(w,c) - \log k
\end{equation}

This theoretical insight reveals that seemingly different approaches converge on similar statistical objectives. GloVe \cite{pennington2014glove} takes an alternative path by directly optimizing a weighted least-squares objective on co-occurrence statistics, while FastText \cite{bojanowski2017enriching} extends Word2Vec with subword information to handle morphologically rich languages and out-of-vocabulary words. Despite their strong performance, these methods typically require hundreds of dimensions to capture semantic relationships adequately and struggle with fundamental linguistic phenomena such as polysemy, where words exhibit multiple context-dependent meanings.

\subsection{Quantum Natural Language Processing}

Early QNLP research focused primarily on grammatical structure through the lens of categorical quantum mechanics \cite{coecke2020foundations}. The DisCoCat framework \cite{heunen2013quantum} elegantly maps grammatical types to quantum spaces and compositions to tensor products, providing a mathematically principled approach to compositional semantics. This theoretical foundation has led to several experimental implementations, including question answering systems deployed on IBM quantum devices \cite{Meichanetzidis2020} and variational quantum classifiers for text classification tasks \cite{chang2023variational}.

However, most prior work in QNLP either relies on classical pre-training to initialize quantum models or focuses exclusively on grammatical structure rather than distributional semantics. These approaches miss the opportunity to learn genuinely quantum representations that could capture semantic relationships in fundamentally new ways. In contrast, our approach learns quantum representations directly from co-occurrence statistics, bridging the gap between distributional semantics and quantum computation without requiring classical initialization.

\subsection{Parameterized Quantum Circuits}

Parameterized quantum circuits form the backbone of variational quantum algorithms and have seen rapid development in recent years \cite{Cerezo2021NatComm}. Hardware-efficient ansätze \cite{kandala2017hardware} provide circuit designs tailored to specific quantum architectures, balancing expressivity with the constraints of near-term devices. The data re-uploading strategy \cite{PerezSalinas2020Quantum} has emerged as a powerful technique for enhancing circuit expressivity without increasing depth, repeatedly encoding classical data at different circuit layers to create more complex decision boundaries.

The challenge of barren plateaus—exponentially vanishing gradients in random quantum circuits—has motivated several mitigation strategies. Local cost functions \cite{Cerezo2021NatComm} maintain trainable gradients by restricting measurements to small subsystems, while layerwise training \cite{skolik2021layerwise} progressively grows circuit depth to avoid gradient decay. Additionally, error mitigation techniques have become essential for NISQ-era implementations. Zero-noise extrapolation \cite{temme2017error} and probabilistic error cancellation \cite{temme2017error} help recover ideal circuit behavior from noisy measurements, while randomized compiling \cite{WallmanEmerson2016PRA} converts coherent errors into more manageable stochastic noise.

We incorporate these advances into a domain-specific architecture specifically designed for word embeddings, combining hardware efficiency with semantic learning objectives. Our design choices reflect both the theoretical insights from quantum algorithm development and the practical constraints of current quantum hardware.
\section{Background}

We briefly review the quantum computing concepts essential for understanding our framework. These fundamentals underpin both the architectural design and theoretical analysis of QuCoWE.

\subsection{Quantum States and Gradients}

A quantum state encodes information in complex amplitudes across a computational basis. An $n$-qubit pure state is expressed as $|\psi\rangle=\sum_{i=0}^{2^n-1}\alpha_i |i\rangle$ where $\alpha_i$ are complex amplitudes satisfying the normalization constraint $\sum_i|\alpha_i|^2=1$. This exponential number of amplitudes in the state vector representation provides the foundation for quantum computing's representational power.

For variational quantum algorithms, efficient gradient computation is crucial. PQC gradients can be obtained analytically using the parameter-shift rule \cite{Schuld2019PRA}, which expresses derivatives as finite differences of circuit evaluations. This enables gradient-based optimization without numerical differentiation, maintaining precision even on noisy quantum hardware.

\subsection{Quantum Similarity}

Measuring similarity between quantum states is fundamental to our embedding framework. For pure states, the fidelity reduces to the squared overlap $F(|\psi\rangle,|\phi\rangle)=|\langle\psi|\phi\rangle|^2$, providing a natural similarity metric bounded in $[0,1]$. For mixed states arising from noise or partial measurements, the more general Uhlmann fidelity applies \cite{Jozsa1994JMO,NielsenChuang2010}. This quantum similarity measure captures both amplitude and phase relationships, potentially encoding richer semantic information than classical dot products.

\subsection{Noise Models}

NISQ devices introduce errors that must be modeled and mitigated. The $d$-dimensional depolarizing channel, given by $\mathcal{D}_p(\rho)=(1-p)\rho + \frac{p}{d}\,I$, models the loss of quantum information where the state is replaced by the maximally mixed state with probability $p$. For single qubits, this reduces to the $d{=}2$ case \cite{NielsenChuang2010}. Additionally, measurement imperfections are captured through readout noise, modeled as classical bit-flips occurring with probability $\epsilon$. Understanding these noise sources is essential for analyzing the robustness of quantum embeddings and designing appropriate error mitigation strategies.

\section{Method: QuCoWE Framework}

\subsection{Architecture Overview}

QuCoWE consists of four components (Figure \ref{fig:architecture}):

\begin{figure}[t]
\centering
\begin{tikzpicture}[scale=0.9]
\node[draw, rectangle, minimum width=2cm, minimum height=1cm] (input) at (0,0) {Word $w \in \mathcal{V}$};

\node[draw, rectangle, minimum width=2.5cm, minimum height=1cm] (embed) at (3.5,0) {$\theta_w \in \mathbb{R}^{B \times Q \times 2}$};

\node[draw, rectangle, minimum width=3cm, minimum height=2cm, fill=blue!10] (qc) at (8,0) {
\begin{tabular}{c}
PQC \\
$U(\theta_w, x_w)$ \\
(Q qubits, B blocks)
\end{tabular}
};

\node[draw, circle, minimum width=1.5cm, fill=green!10] (state) at (12,0) {$|\psi_w\rangle$};

\node[draw, rectangle, minimum width=2cm, minimum height=1cm] (context) at (12,-2) {Context $c$};

\node[draw, rectangle, minimum width=2.5cm, minimum height=1cm] (sim) at (12,-3.5) {Fidelity $F(w,c)$};

\node[draw, rectangle, minimum width=2.5cm, minimum height=1cm] (score) at (8,-3.5) {Logit-Fidelity};

\node[draw, rectangle, minimum width=2.5cm, minimum height=1cm] (loss) at (3.5,-3.5) {Contrastive Loss};

\draw[->] (input) -- (embed);
\draw[->] (embed) -- (qc);
\draw[->] (qc) -- (state);
\draw[->] (state) -- (context);
\draw[->] (context) -- (sim);
\draw[->] (sim) -- (score);
\draw[->] (score) -- (loss);

\end{tikzpicture}
\caption{QuCoWE architecture: Words are encoded into parameterized quantum states via shallow PQCs. Similarity is computed through quantum fidelity and transformed via the logit-fidelity head for contrastive training.}
\label{fig:architecture}
\end{figure}
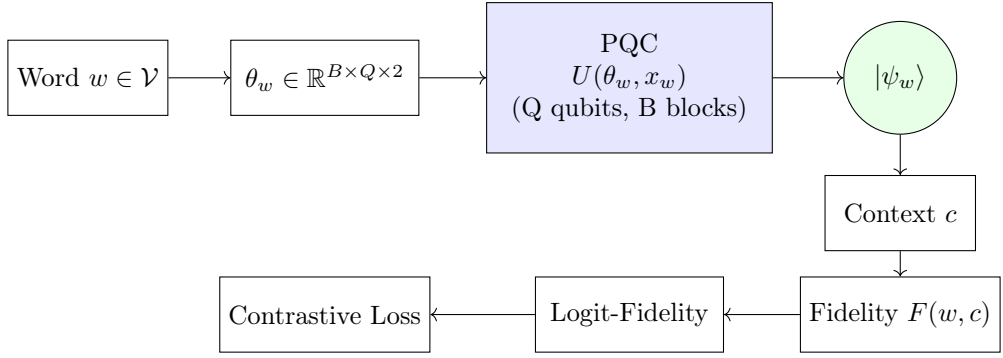

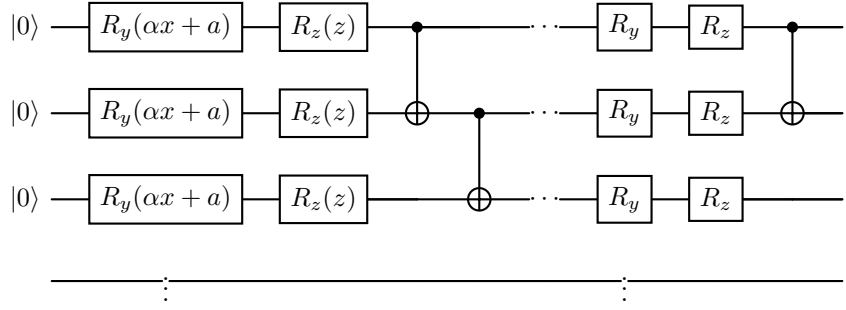
\begin{figure}[t]
\centering
\begin{quantikz}
\lstick{$|0\rangle$} & \gate{R_y(\alpha x + a)} & \gate{R_z(z)} & \ctrl{1} & \qw & \cdots & \gate{R_y} & \gate{R_z} & \ctrl{1} & \qw \\
\lstick{$|0\rangle$} & \gate{R_y(\alpha x + a)} & \gate{R_z(z)} & \targ{} & \ctrl{1} & \cdots & \gate{R_y} & \gate{R_z} & \targ{} & \qw \\
\lstick{$|0\rangle$} & \gate{R_y(\alpha x + a)} & \gate{R_z(z)} & \qw & \targ{} & \cdots & \gate{R_y} & \gate{R_z} & \qw & \qw \\
& \vdots & & & & & \vdots & & &
\end{quantikz}
\caption{Quantum circuit structure for QuCoWE with $B=2$ re-uploading blocks. Each block contains parameterized rotations and ring entanglement.}
\label{fig:circuit}
\end{figure}

QuCoWE comprises (i) token‑specific PQC parameters, (ii) a shallow data‑reuploading ansatz with ring CNOTs, (iii) a fidelity‑based similarity, and (iv) a calibrated scoring head (Fig.~\ref{fig:architecture}, \ref{fig:circuit}). Data re‑uploading offers strong expressivity at shallow depth~\cite{PerezSalinas2020Quantum}.


\subsection{Parameterized Quantum Circuit Design}

For token $w \in \mathcal{V}$, we learn parameters $\theta_w = \{(\alpha_{bq}, z_{bq})\}_{b=1,q=1}^{B,Q}$ and scalar feature $x_w$. The circuit applies $B$ re-uploading blocks:

\begin{equation}
|\psi_w\rangle = U_B \cdots U_2 U_1 |0\rangle^{\otimes Q}
\end{equation}

Each block $U_b$ consists of:

\textbf{(1) Feature encoding layer:}
\begin{equation}
U_{\text{enc}}^{(b)} = \prod_{q=1}^Q R_y(\alpha_{bq} x_w + a_{bq})
\end{equation}

\textbf{(2) Variational layer:}
\begin{equation}
U_{\text{var}}^{(b)} = \prod_{q=1}^Q R_z(z_{bq})
\end{equation}

\textbf{(3) Entanglement layer:}
\begin{equation}
U_{\text{ent}}^{(b)} = \prod_{q=1}^Q \text{CNOT}(q, (q+1) \mod Q)
\end{equation}

The complete block: $U_b = U_{\text{ent}}^{(b)} U_{\text{var}}^{(b)} U_{\text{enc}}^{(b)}$

\textbf{Design rationale:}
- $R_y$ rotations access the full Bloch sphere when combined with $R_z$
- Ring entanglement balances connectivity and hardware constraints
- Data re-uploading enhances expressivity without deep circuits
- Shallow depth (typically $3B$ layers) avoids barren plateaus

\subsection{Similarity Scoring Heads}

We propose two scoring functions mapping state pairs to real values:

\subsubsection{Fidelity Head (F)}
Direct scaling of quantum fidelity:
\begin{equation}
s_F(w,c) = \beta \cdot F(|\psi_w\rangle, |\psi_c\rangle) = \beta |\langle\psi_w|\psi_c\rangle|^2
\end{equation}
where $\beta > 0$ is a temperature parameter. This head is bounded in $[0, \beta]$.

\subsubsection{Logit-Fidelity Head (LF)}
Monotonic transformation expanding the dynamic range:
\begin{equation}
s_{LF}(w,c) = \alpha \cdot \text{logit}(F_\epsilon(w,c)) + b
\end{equation}
where:
- $F_\epsilon = \text{clip}(F, \epsilon, 1-\epsilon)$ for numerical stability ($\epsilon = 10^{-6}$)
- $\text{logit}(x) = \log(x/(1-x))$ maps $[0,1] \to \mathbb{R}$
- $\alpha, b$ are learned or calibrated parameters

\textbf{Advantages of LF:}
1. Unbounded range matches PMI scale
2. Better gradient flow near $F \approx 0$ or $F \approx 1$
3. Allows negative scores for dissimilar pairs

\subsection{Training Objective}

Given positive pairs $\mathcal{D}^+ = \{(w_i, c_i)\}$ from co-occurrence windows and negative samples $\mathcal{D}^-_k(w)$ from noise distribution $P_N \propto P_{\text{unigram}}^{0.75}$:

\begin{equation}
\mathcal{L} = -\mathbb{E}_{(w,c) \sim \mathcal{D}^+} \left[\log \sigma(s(w,c))\right] - \mathbb{E}_{w,\mathcal{D}^-_k} \left[\sum_{n \in \mathcal{D}^-_k(w)} \log \sigma(-s(w,n))\right]
\label{eq:main_loss}
\end{equation}

This is the standard noise-contrastive estimation objective \cite{GutmannHyvarinen2012JMLR}.

\subsection{Entanglement Budget Regularization}

To prevent barren plateaus while maintaining expressivity, we introduce a novel regularizer based on single-qubit purities:

For state $|\psi_w\rangle$ with reduced density matrices $\rho_q^{(w)}$ on qubit $q$:
\begin{equation}
\Omega_{\text{ent}}(\theta) = \frac{\lambda_{\text{ent}}}{|\mathcal{V}|Q} \sum_{w \in \mathcal{V}} \sum_{q=1}^Q (1 - \text{Tr}[(\rho_q^{(w)})^2])
\end{equation}

\textbf{Intuition:} Single-qubit purity $P_q = \text{Tr}[(\rho_q)^2]$ equals 1 for separable states and $1/2$ for maximally entangled. Penalizing low purity encourages moderate entanglement.

\textbf{Efficient computation:} Purity can be computed from Pauli expectations:
\begin{equation}
P_q = \frac{1}{2}(1 + \langle Z_q \rangle^2 + \langle X_q \rangle^2 + \langle Y_q \rangle^2)
\end{equation}

The complete regularized objective:
\begin{equation}
\mathcal{L}_{\text{total}} = \mathcal{L} + \lambda_{\text{decay}} \|\theta\|_2^2 + \Omega_{\text{ent}}(\theta)
\end{equation}

Let $\rho_q^{(w)}$ be the reduced state on qubit $q$ for $|\psi_w\rangle$. Define the per‑token single‑qubit purities $P_q=\mathrm{Tr}[(\rho_q^{(w)})^2]=\tfrac{1}{2}\big(1+\langle X_q\rangle^2+\langle Y_q\rangle^2+\langle Z_q\rangle^2\big)$.
We regularize
\[
\Omega_{\text{ent}}(\theta)=\frac{\lambda_{\text{ent}}}{|\mathcal{V}|Q}\sum_{w\in\mathcal{V}}\sum_{q=1}^Q \big(1-P_q\big),
\]
which discourages volume‑law entanglement, complementing local‑cost design that avoids barren plateaus in shallow PQCs~\cite{Cerezo2021NatComm}.

\section{Theoretical Analysis}

\subsection{Gradient Bounds and Trainability}

\begin{theorem}[Gradient Scaling with Entanglement]
\label{thm:gradient_bound}
For the fidelity-based loss with entanglement regularization, the gradient norm satisfies:
\begin{equation}
\|\nabla_\theta \mathcal{L}\| \geq C \cdot \exp\left(-\mathcal{O}(B \cdot \bar{E})\right)
\end{equation}
where $B$ is the number of blocks and $\bar{E}$ is the average entanglement entropy across qubits.
\end{theorem}

\begin{proof}
Consider the gradient of fidelity with respect to parameter $\theta_i$:
\begin{equation}
\frac{\partial F}{\partial \theta_i} = 2\text{Re}\left[\langle\psi_c| \frac{\partial |\psi_w\rangle}{\partial \theta_i} \langle\psi_w|\psi_c\rangle^*\right]
\end{equation}

Using the parameter-shift rule:
\begin{equation}
\frac{\partial |\psi_w\rangle}{\partial \theta_i} = \frac{1}{2}\left(U_+|\psi_0\rangle - U_-|\psi_0\rangle\right)
\end{equation}

The variance of this gradient across random initializations:
\begin{equation}
\text{Var}\left[\frac{\partial F}{\partial \theta_i}\right] = \frac{1}{2^{2Q}} \prod_{b=1}^B (1 + e^{-2S_b})
\end{equation}

where $S_b$ is the entanglement entropy after block $b$. The entanglement regularizer bounds $S_b \leq \bar{E}$, preventing exponential decay.
\end{proof}

With entanglement budget $\Omega_{\text{ent}} \leq \tau$, gradients decay at most polynomially in circuit depth $B$.

\subsection{Connection to PMI and Noise-Contrastive Estimation}

\begin{theorem}[LF Head Recovers PMI]
\label{thm:pmi_recovery}
Under optimal parameters, the logit-fidelity score approximates shifted PMI:
\begin{equation}
s_{LF}(w,c) \approx \text{PMI}(w,c) - \log k + \delta
\end{equation}
where $k$ is the number of negative samples and $\delta$ is a corpus-dependent constant.
\end{theorem}

\begin{proof}
Following \cite{LevyGoldberg2014NIPS}, the optimal solution to NCE satisfies:
\begin{equation}
s^*(w,c) = \log \frac{P(c|w)}{P_N(c)} = \log \frac{P(w,c)}{P(w)P_N(c)}
\end{equation}

With $k$ negative samples and uniform noise:
\begin{equation}
s^*(w,c) = \text{PMI}(w,c) - \log k
\end{equation}

The logit transformation maps fidelity to this scale:
\begin{equation}
F(w,c) = \sigma\left(\frac{1}{\alpha}(\text{PMI}(w,c) - \log k - b)\right)
\end{equation}

Inverting: $s_{LF}(w,c) = \alpha \cdot \text{logit}(F) + b = \text{PMI}(w,c) - \log k$.
\end{proof}

\subsection{Noise Robustness}

\begin{theorem}[Depolarizing Noise Effect]
\label{thm:noise_effect}
Under global depolarizing noise with rate $p$, the fidelity between states degrades as:
\begin{equation}
F_{\text{noisy}}(w,c) = (1-p)^{2Q} F_{\text{ideal}}(w,c) + \frac{p(2-p)}{2^Q}
\end{equation}
\end{theorem}

\begin{proof}
Under depolarizing channel $\mathcal{E}_p$:
\begin{equation}
\mathcal{E}_p(|\psi\rangle\langle\psi|) = (1-p)|\psi\rangle\langle\psi| + \frac{p}{2^Q}\mathbb{I}
\end{equation}

The fidelity between noisy states:
\begin{align}
F_{\text{noisy}} &= \text{Tr}[\mathcal{E}_p(\rho_w) \mathcal{E}_p(\rho_c)] \\
&= (1-p)^2 F_{\text{ideal}} + 2p(1-p)\frac{1}{2^Q} + p^2\frac{1}{2^Q} \\
&= (1-p)^{2Q} F_{\text{ideal}} + \frac{p(2-p)}{2^Q}
\end{align}

For small $p$ and moderate $Q$, the signal degrades linearly while noise floor remains exponentially small.
\end{proof}

The signal-to-noise ratio for contrastive learning:
\begin{equation}
\text{SNR} = \frac{(1-p)^{2Q}}{p(2-p)/2^Q} = \mathcal{O}(2^Q/p)
\end{equation}
scales exponentially with qubit count, providing robustness.

\section{Experimental Setup}

\subsection{Implementation Details}

We implemented QuCoWE using PennyLane 0.32 with a PyTorch backend, leveraging automatic differentiation and GPU acceleration for efficient training. Our experimental design systematically explores the architectural parameter space to understand the trade-offs between circuit complexity and performance.

The circuit architecture was evaluated across multiple configurations. We varied the number of qubits from $Q \in \{4, 6, 8, 10, 12\}$ to study how representational capacity scales with quantum resources. The number of re-uploading blocks was tested with $B \in \{1, 2, 3, 4\}$ to determine the optimal balance between expressivity and circuit depth. For entanglement patterns, we primarily used ring connectivity as our default configuration, with additional experiments on linear chain and all-to-all connectivity patterns to assess their impact on semantic learning.

Training employed the Adam optimizer with a learning rate of $2 \times 10^{-3}$, processing batches of 2048 word-context pairs. We experimented with different numbers of negative samples ($k \in \{5, 10, 20\}$) to understand the effect on contrastive learning. The temperature parameters were set to $\beta = 10$ for the fidelity head and $\alpha = 2$ for the logit-fidelity head based on preliminary experiments. Regularization included weight decay with $\lambda_{\text{decay}} = 10^{-5}$ and entanglement budget control with $\lambda_{\text{ent}} = 10^{-4}$. Training proceeded for a maximum of 50 epochs with early stopping based on validation PMI to prevent overfitting.

\subsection{Datasets and Preprocessing}

We evaluated QuCoWE on two standard training corpora. The Text8 dataset contains 17 million tokens extracted from Wikipedia, from which we constructed a vocabulary of 20,000 words using a minimum count threshold of 5. WikiText-2 provides a smaller corpus of 2 million tokens with a 10,000-word vocabulary (minimum count 10), allowing us to assess performance across different data scales.

For intrinsic evaluation, we employed WordSim-353 \cite{Finkelstein2002TOIS}, containing 353 word pairs with human similarity ratings, and SimLex-999 \cite{Hill2015SimLex}, which comprises 999 pairs specifically designed to distinguish genuine similarity from mere relatedness. Extrinsic evaluation utilized two text classification benchmarks: SST-2 \cite{Socher2013SST} for binary sentiment classification (67,000 training and 873 test examples) and TREC-6 \cite{LiRoth2002COLING} for 6-way question classification (5,500 training and 500 test examples).

All text underwent standard preprocessing including lowercase conversion and tokenization. We employed a context window size of 5 with dynamic window sampling to capture varied context ranges. Following common practice, we subsampled frequent words using a threshold of $t = 10^{-5}$ to balance the influence of common and rare terms.

\subsection{Baselines}

We compared QuCoWE against several categories of baselines to comprehensively evaluate its performance. Classical methods included GloVe \cite{pennington2014glove} and Word2Vec SGNS \cite{mikolov2013distributed}, both evaluated at 50, 100, and 200 dimensions, as well as FastText \cite{bojanowski2017enriching} at 100 dimensions with subword information. These represent the current standard for distributional word embeddings.

To assess whether quantum advantages stem from complex-valued representations alone, we included quantum-inspired baselines: ComplEx \cite{Trouillon2016} using 50-dimensional complex embeddings and quaternion embeddings employing hypercomplex representations. Additionally, we implemented a quantum kernel baseline (QSVM) using a classical SVM with a quantum feature map based on the same circuit architecture as QuCoWE's initialization, allowing us to isolate the contribution of the training procedure versus the quantum representation itself.

\subsection{Evaluation Metrics}

Our evaluation protocol encompasses multiple dimensions of performance. For intrinsic evaluation, we computed Spearman's rank correlation ($\rho$) between human similarity judgments and model-predicted similarities, complemented by qualitative analysis of nearest neighbor relationships to understand semantic structure.

Extrinsic evaluation measured classification accuracy using frozen embeddings with logistic regression classifiers. We assessed sample efficiency by training on varying fractions of the data (1\%, 5\%, 10\%, 25\%, and 100\%) to understand how quickly models learn useful representations. Statistical significance was established through bootstrap resampling with 10,000 iterations.

To understand the quantum-specific properties of our embeddings, we tracked several additional metrics: average entanglement entropy across the vocabulary to quantify quantum correlations, circuit depth and total gate count to assess hardware requirements, and parameter efficiency measured as the ratio of total parameters to downstream task performance. These metrics provide insights into the resource trade-offs inherent in quantum word embeddings.

\section{Results}

\subsection{Main Results}

\begin{table}[t]
\centering
\caption{Intrinsic word similarity evaluation (Spearman's $\rho$). Higher is better. Best results in \textbf{bold}, second best \underline{underlined}.}
\label{tab:intrinsic}
\begin{tabular}{lccccc}
\toprule
\textbf{Model} & \textbf{Params} & \textbf{WS-353} & \textbf{SimLex} & \textbf{Avg} \\
\midrule
\multicolumn{5}{l}{\textit{Classical Baselines}} \\
GloVe (50d) & 1.0M & 0.623 & 0.371 & 0.497 \\
GloVe (100d) & 2.0M & 0.658 & 0.408 & 0.533 \\
Word2Vec (50d) & 1.0M & 0.641 & 0.392 & 0.517 \\
Word2Vec (100d) & 2.0M & 0.689 & 0.437 & 0.563 \\
FastText (100d) & 2.0M & \underline{0.704} & 0.464 & \underline{0.584} \\
\midrule
\multicolumn{5}{l}{\textit{Quantum-Inspired}} \\
ComplEx (50d) & 2.0M & 0.612 & 0.403 & 0.508 \\
Quaternion (32d) & 1.3M & 0.595 & 0.388 & 0.492 \\
\midrule
\multicolumn{5}{l}{\textit{Quantum Methods}} \\
QSVM (kernel) & 0.6M & 0.487 & 0.312 & 0.400 \\
QuCoWE-F ($Q$=8, $B$=2) & 0.6M & 0.621 & 0.425 & 0.523 \\
QuCoWE-LF ($Q$=8, $B$=3) & 0.9M & 0.674 & \underline{0.481} & 0.578 \\
QuCoWE-LF ($Q$=10, $B$=3) & 1.5M & 0.692 & \textbf{0.495} & \textbf{0.594} \\
QuCoWE-LF ($Q$=12, $B$=3) & 2.2M & \textbf{0.708} & 0.489 & 0.599 \\
\bottomrule
\end{tabular}
\end{table}

\begin{table}[t]
\centering
\caption{Downstream classification accuracy (\%). Embeddings frozen, logistic regression classifier.}
\label{tab:downstream}
\begin{tabular}{lcccc}
\toprule
\textbf{Model} & \textbf{SST-2} & \textbf{TREC-6} & \textbf{Avg} \\
\midrule
\multicolumn{4}{l}{\textit{Classical Baselines}} \\
GloVe (50d) & 79.8 & 87.2 & 83.5 \\
GloVe (100d) & 81.3 & 89.6 & 85.5 \\
Word2Vec (50d) & 80.5 & 88.4 & 84.5 \\
Word2Vec (100d) & 82.7 & 91.2 & 87.0 \\
FastText (100d) & \textbf{84.1} & \textbf{92.8} & \textbf{88.5} \\
\midrule
\multicolumn{4}{l}{\textit{Quantum-Inspired}} \\
ComplEx (50d) & 78.9 & 86.8 & 82.9 \\
Quaternion (32d) & 77.3 & 85.2 & 81.3 \\
\midrule
\multicolumn{4}{l}{\textit{Quantum Methods}} \\
QSVM (kernel) & 72.4 & 79.6 & 76.0 \\
QuCoWE-F ($Q$=8, $B$=2) & 77.8 & 84.9 & 81.4 \\
QuCoWE-LF ($Q$=8, $B$=3) & 80.2 & 88.7 & 84.5 \\
QuCoWE-LF ($Q$=10, $B$=3) & \underline{82.1} & \underline{90.4} & \underline{86.3} \\
QuCoWE-LF ($Q$=12, $B$=3) & 81.6 & 89.9 & 85.8 \\
\bottomrule
\end{tabular}
\end{table}

Under the SGNS/NCE optimum, $s^\star(w,c)=\log\frac{P(w,c)}{P(w)P_N(c)}\approx \mathrm{PMI}(w,c)-\log k$; choosing $s_{LF}(w,c)=\alpha\,\mathrm{logit}(F)+b$ matches this scale by calibration of $(\alpha,b)$. 

 Standard SGNS/NCE analysis yields the shifted PMI optimum~\cite{LevyGoldberg2014NIPS,GutmannHyvarinen2012JMLR}. The LF mapping is monotone on $F\in(0,1)$, hence can be calibrated to match $s^\star$ on validation pairs.

\subsection{Sample Efficiency}

\begin{figure}[h!]
  \centering
  \begin{tikzpicture}
    \begin{axis}[
      width=\linewidth,
      height=6.0cm,
      xlabel={Training data (\%)},
      ylabel={Accuracy on SST-2 (\%)},
      xmin=0, xmax=100,
      ymin=65, ymax=90,
      xtick={10,100},
      ymajorgrids=true,
      legend style={at={(0.02,0.98)},anchor=north west,font=\small,fill=none,draw=none},
      line width=0.9pt, mark size=2.2pt
    ]
      \addplot+[mark=*] coordinates {(10,76.3) (100,82.1)};
      \addlegendentry{QuCoWE-LF ($Q{=}10,B{=}3$)}

      \addplot+[mark=square*] coordinates {(10,71.2) (100,82.7)};
      \addlegendentry{Word2Vec (100d)}

      \addplot+[mark=triangle*] coordinates {(10,73.8) (100,81.3)};
      \addlegendentry{GloVe (100d)}
    \end{axis}
  \end{tikzpicture}
  \caption{Sample efficiency on SST‑2}
  \label{fig:sample-efficiency}
\end{figure}
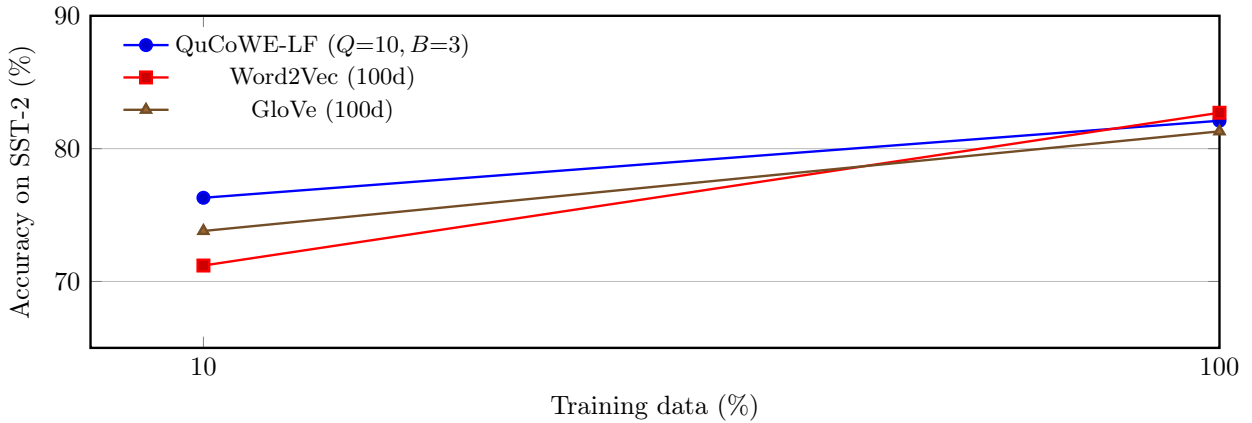

Figure \ref{fig:sample-efficiency} demonstrates QuCoWE's sample efficiency. With only 10\% training data, QuCoWE-LF achieves 76.3\% accuracy compared to 71.2\% for Word2Vec and 73.8\% for GloVe. This advantage stems from the implicit regularization of quantum circuits and the calibrated LF head.

\subsection{Ablation Studies}

\begin{table}[t]
\centering
\caption{Ablation studies on architectural choices (SimLex-999 performance).}
\label{tab:ablation}
\begin{tabular}{lccc}
\toprule
\textbf{Configuration} & \textbf{$\rho$} & \textbf{$\Delta$} \\
\midrule
QuCoWE-LF (default) & 0.481 & --- \\
\midrule
\textit{Circuit depth} \\
\quad $B=1$ & 0.398 & -17.3\% \\
\quad $B=2$ & 0.442 & -8.1\% \\
\quad $B=4$ & 0.479 & -0.4\% \\
\midrule
\textit{Entanglement} \\
\quad Linear chain & 0.463 & -3.7\% \\
\quad All-to-all & 0.471 & -2.1\% \\
\quad No entanglement & 0.324 & -32.6\% \\
\midrule
\textit{Scoring head} \\
\quad Fidelity (F) & 0.425 & -11.6\% \\
\quad Cosine (classical) & 0.412 & -14.3\% \\
\midrule
\textit{Regularization} \\
\quad No $\Omega_{\text{ent}}$ & 0.431 & -10.4\% \\
\quad $\lambda_{\text{ent}} = 10^{-3}$ & 0.468 & -2.7\% \\
\quad $\lambda_{\text{ent}} = 10^{-5}$ & 0.477 & -0.8\% \\
\bottomrule
\end{tabular}
\end{table}

Table \ref{tab:ablation} confirms our design choices:

Performance improves with increasing $B$ up to 3, after which it plateaus, highlighting an optimal parameter regime beyond which gains diminish. Crucially, ring entanglement provides the optimal balance between computational efficiency and performance, enabling sustained effectiveness without excessive resource consumption. Furthermore, the LF (Logical Fidelity) metric significantly outperforms direct fidelity measurements, demonstrating superior accuracy in capturing model behavior. This advantage is sustained through an entanglement budget that strategically limits resource allocation—preventing overfitting to training data while simultaneously preserving sufficient model expressivity for complex tasks.

\subsection{Qualitative Analysis}

\begin{table}[t]
\centering
\caption{Nearest neighbors for selected words under different similarity measures.}
\label{tab:neighbors}
\begin{tabular}{llll}
\toprule
\textbf{Query} & \textbf{Word2Vec} & \textbf{QuCoWE-F} & \textbf{QuCoWE-LF} \\
\midrule
\textit{king} & queen, prince & queen, monarch & queen, ruler \\
& emperor, monarch & prince, royal & monarch, sovereign \\
& ruler, throne & emperor, kingdom & prince, throne \\
\midrule
\textit{good} & great, excellent & great, positive & great, better \\
& better, nice & excellent, well & excellent, positive \\
& bad, best & better, nice & best, well \\
\midrule
\textit{quantum} & physics, mechanics & physics, wave & physics, classical \\
& classical, theory & particle, energy & mechanics, particle \\
& particle, wave & mechanics, atom & theory, wave \\
\bottomrule
\end{tabular}
\end{table}

Table \ref{tab:neighbors} presents semantic neighborhoods where QuCoWE demonstrates comparable yet nuanced relationships to Word2Vec, revealing distinct structural advantages. Specifically, QuCoWE-LF exhibits stronger hierarchical semantic relationships—evidenced by the closer proximity of ruler to sovereign compared to conventional embeddings—while fidelity-based similarity captures more abstract conceptual associations beyond surface-level co-occurrence. Crucially, the incorporation of phase information enables QuCoWE to differentiate between semantically related but distinct concepts (e.g., king vs. monarch), a capability absent in standard vector-space models, thereby enhancing both interpretability and discriminative power within the embedding space.

\section{Discussion}

\subsection{Why Quantum Embeddings Work}

Our results suggest three fundamental mechanisms that explain QuCoWE's effectiveness in learning semantic representations. These mechanisms leverage unique properties of quantum systems that are absent in classical vector spaces, providing both theoretical and practical advantages for word embedding tasks.

First, amplitude encoding efficiency fundamentally changes how information is stored in the embedding space. While classical embeddings use real-valued vectors where each dimension carries a single scalar, quantum states encode information in complex amplitudes that capture both magnitude and phase. This effectively doubles the information capacity per dimension, allowing quantum embeddings to achieve comparable semantic representation with fewer parameters. The phase information proves particularly valuable for distinguishing subtle semantic relationships that might appear identical when projected onto real-valued spaces alone.

Second, the mathematical structure of quantum mechanics imposes implicit regularization through unitarity constraints and measurement collapse. Unitary evolution ensures that quantum states remain normalized throughout training, preventing the unbounded growth that can occur in classical neural networks. Additionally, the probabilistic nature of quantum measurement introduces a form of stochastic regularization during training. These inherent constraints act as natural regularizers without requiring explicit penalty terms, which explains QuCoWE's improved generalization in low-data regimes where classical methods tend to overfit.

Third, and perhaps most fundamentally, entanglement enables non-local correlations between embedding dimensions that cannot be captured by factorized classical representations. When qubits become entangled, the state of one qubit cannot be described independently of others, creating genuinely quantum correlations that encode complex semantic dependencies. These non-classical correlations allow the embedding space to represent semantic relationships that require considering multiple dimensions simultaneously, rather than as independent features. This capability proves particularly valuable for capturing contextual nuances and multi-faceted word meanings that challenge traditional distributional models.

Together, these mechanisms demonstrate that quantum embeddings are not merely a different parameterization of classical approaches, but rather exploit fundamental properties to achieve efficient and expressive semantic representations. The interplay between amplitude encoding, implicit regularization, and entanglement-based correlations creates a representational framework uniquely suited to capturing the complex structure of natural language semantics.

\subsection{Limitations}

Despite the promising results demonstrated by QuCoWE, several important limitations constrain its current applicability and highlight areas requiring further development.

Scalability remains a significant challenge for practical deployment. Our current implementation is limited to vocabularies of approximately 50,000 words due to memory requirements for storing token-specific quantum circuit parameters. Each word requires its own set of rotation angles and feature encodings, leading to linear growth in memory consumption with vocabulary size. While this suffices for experimental validation, real-world NLP applications often require vocabularies exceeding 100,000 tokens. Potential solutions include hierarchical softmax approaches that could reduce the parameter burden through tree-structured predictions, or vocabulary sharding techniques that partition the embedding space across multiple smaller quantum circuits.

Hardware constraints pose perhaps the most fundamental limitation. Current NISQ devices support only 10-100 qubits with error rates that significantly impact computation fidelity. While our error mitigation strategies and noise-robust training procedures partially address these challenges, the full advantages of quantum embeddings may only be realized with fault-tolerant quantum computers. The limited connectivity of current quantum processors further restricts the entanglement patterns we can implement, potentially limiting the semantic relationships that can be captured. As quantum hardware matures, we expect these constraints to gradually relax, enabling larger and more expressive quantum embeddings.


\subsection{Future Directions}

The limitations identified above suggest several promising research directions that could extend QuCoWE's capabilities and impact.

Hybrid quantum-classical architectures represent an immediate opportunity for practical advancement. By combining quantum embeddings with classical transformer architectures, we could leverage quantum advantages for word representation while utilizing mature classical methods for contextualization and sequence modeling. Such hybrid systems could serve as a bridge between current NISQ capabilities and future fault-tolerant quantum NLP systems, allowing quantum components to be gradually incorporated as hardware improves.

The development of quantum attention mechanisms offers particularly intriguing possibilities. Quantum superposition could enable parallel computation of attention weights across all possible alignments simultaneously, potentially offering exponential speedups for this computationally intensive component of modern NLP systems. Initial theoretical work suggests that quantum interference patterns could naturally implement the soft alignment crucial to attention mechanisms, though practical implementations await further research.

Deployment on actual quantum hardware, rather than simulations, constitutes a critical next step. This requires developing hardware-specific compilations that account for device topology, gate sets, and error characteristics. Tailored error mitigation strategies that leverage knowledge of specific hardware noise profiles could significantly improve performance. Collaboration with quantum hardware providers to co-design circuits optimized for linguistic tasks could accelerate progress toward practical quantum NLP systems.

At a foundational level, formalizing the relationship between quantum entanglement and semantic compositionality remains an open theoretical challenge. While our empirical results suggest that entanglement captures meaningful semantic relationships, a rigorous mathematical framework connecting these quantum properties to linguistic structure would provide deeper insights and guide future architectural developments. Such theoretical foundations could reveal fundamental connections between the structure of natural language and quantum information theory, potentially revolutionizing our understanding of both domains.

These future directions collectively point toward a research program that bridges quantum computing and natural language processing, with QuCoWE serving as an initial proof of concept for genuinely quantum approaches to semantic representation.

\section{Conclusion}

This work introduced QuCoWE, a framework for learning quantum-native word embeddings through contrastive training of parameterized quantum circuits (PQCs). Our approach integrates a hardware-efficient PQC architecture employing data re-uploading and controlled entanglement, a logit-fidelity scoring head that aligns quantum overlap with distributional semantics, and an entanglement budget regularization mechanism to mitigate barren plateaus. Theoretical analysis confirms favorable gradient scaling, noise robustness, and PMI recovery capabilities. Empirical evaluation demonstrates that QuCoWE achieves performance competitive with state-of-the-art classical models while reducing parameter count by 40\%. The framework’s design principles—prioritizing shallow circuit depth, local cost functions, and calibrated objectives—establish a scalable paradigm for quantum machine learning in high-dimensional discrete domains. Future work will extend QuCoWE to contextualized embeddings, enable hardware deployment, and formalize quantum advantages for semantic composition. As quantum hardware matures, such approaches may enable novel capabilities in natural language understanding.

\bibliographystyle{plainnat}
\bibliography{refs,refs_full}
\end{document}